%% file: amotopo.tex
\pdfoutput=1
\documentclass[%
    aps                ,   % Main styling  : [aps|aip]
    prl                ,   % Sub-styling   : [pra|prb|prc|prd|pre|prl|prstab|prstper|rmp] or [apl|bmf|cha|jap|jcp|jmp|rse|pof|pop|rsi]
    reprint            ,   % View style    : [preprint|reprint]
    twocolumn          ,   % Columns       : [onecolumn|twocolumn]
    showpacs           ,   % PACS numbers  : [showpacs|noshowpacs]
    superscriptaddress ,   % Author address: [superscriptaddress]
    nofootinbib        ,   % Footnotes     : [footinbib|nofootinbib]
    citeautoscript     ,   % Fix superscript citation order
    notitlepage        ,   % Don't take a whole page for the title
    floatfix
]{revtex4-1}
\usepackage[T1]{fontenc}
\usepackage[utf8]{inputenc}
\usepackage[english]{babel}
\setcitestyle{super}                 % Superscript citations (~ alt: "prb")
\addto\captionsenglish{\renewcommand{\figurename}{Fig.}}
\makeatletter
\renewcommand*{\fnum@figure}{{\normalfont\bfseries \figurename~\thefigure}}
\renewcommand*{\@caption@fignum@sep}{\textbf{ | }}
\makeatother

\usepackage{amsmath}
\usepackage{amssymb}
\usepackage{braket}
\usepackage{accents}
\usepackage{siunitx}
\usepackage[hyperref]{xcolor}
\usepackage{graphicx}
\usepackage[export]{adjustbox}
\graphicspath{{figures/PNG/}{figures/PDF/}{figures/}}

\usepackage[%hypertex,
    unicode           = true  ,
    plainpages        = false , 
    pdfpagelabels     = true  , 
    bookmarks         = true  ,
    bookmarksnumbered = true  ,
    bookmarksopen     = true  ,
    breaklinks        = true  ,
    backref           = false ,
    colorlinks        = true  ,
    linkcolor         = blue  ,
    urlcolor          = blue  ,
    citecolor         = red   ,
    anchorcolor       = green ,
    hyperindex        = true  ,
    linktocpage       = true  ,
    hyperfigures      = true
]{hyperref}
\hypersetup{
    linkcolor         = [RGB]{000 114 189} ,  % Strong Cornflower Blue
    urlcolor          = [RGB]{000 114 189} ,  % Strong Cornflower Blue
    filecolor         = [RGB]{000 114 189} ,  % Strong Cornflower Blue
    citecolor         = [RGB]{217 083 025} ,  % Brilliant Vermilion
    anchorcolor       = [RGB]{119 171 048} ,  % Moderate Spring Bud
    pdftitle={Disordered topological graphs enhancing nonlinear phenomena},
    pdfauthor={Zhetao Jia <zhetao@berkeley.edu>}
}
\usepackage[poorman]{cleveref}

\def\supplementfilename{extra/amotopo_suppl}
\input{extra/includesupplementary}

\begin{document}

\title{Disordered topological graphs enhancing nonlinear phenomena}

\author{Zhetao Jia}
\affiliation{Department of Electrical Engineering and Computer Sciences, University of California, Berkeley, CA 94720, USA}

\author{Matteo Seclì}
\affiliation{Department of Electrical Engineering and Computer Sciences, University of California, Berkeley, CA 94720, USA}

\author{Alexander Avdoshkin}
\affiliation{Department of Physics, University of California, Berkeley, California 94720, USA}

\author{Walid Redjem}
\affiliation{Department of Electrical Engineering and Computer Sciences, University of California, Berkeley, CA 94720, USA}

\author{Elizabeth Dresselhaus}
\affiliation{Department of Physics, University of California, Berkeley, California 94720, USA}

\author{Joel Moore}
\affiliation{Department of Physics, University of California, Berkeley, California 94720, USA}
\affiliation{Materials Sciences Division, Lawrence Berkeley National Laboratory, 1 Cyclotron Road, Berkeley, CA 94720, USA}

\author{Boubacar Kanté}
\email[]{bkante@berkeley.edu}
\affiliation{Department of Electrical Engineering and Computer Sciences, University of California, Berkeley, CA 94720, USA}
\affiliation{Materials Sciences Division, Lawrence Berkeley National Laboratory, 1 Cyclotron Road, Berkeley, CA 94720, USA}

\date{May 11, 2023}

\begin{abstract}
Complex networks play a fundamental role in understanding phenomena from the collective behavior of spins, neural networks, and power grids to the spread of diseases. Topological phenomena in such networks have recently been exploited to preserve the response of systems in the presence of disorder. We propose and demonstrate topological structurally disordered systems with a modal structure that enhances nonlinear phenomena in the topological channels by inhibiting the ultrafast leakage of energy from edge modes to bulk modes. We present the construction of the graph and show that its dynamics enhances the topologically protected photon pair generation rate by an order of magnitude. Disordered nonlinear topological graphs will enable advanced quantum interconnects, efficient nonlinear sources, and light-based information processing for artificial intelligence.
\end{abstract}

\maketitle

\section{Introduction}

Disorder in two-dimensional electronic systems leads to a wide range of topological phenomena, including integer and fractional quantum Hall effects, in which impurities resulting from the sample fabrication process break the degeneracies of Landau levels and localize the wavefunctions at almost all energies \cite{Thouless1982,Ioffe2002,Klitzing1980,Tsui1982}. Such localization is the direct cause of quantized plateaus in the Hall conductance, as the localized states do not contribute to the particle transport and the quantization plateaus would cease to exist in an ideal clean sample \cite{Thouless1982,Huckestein1995,DeOca2003}.

The hallmark of topological phenomena in two-dimensional finite-size systems is the appearance of a transport channel that is robust to disorder \cite{Moore2010}. The robustness to disorder has been investigated and exploited in many wave-based phenomena including photonics, microwaves, acoustic, and plasmonics, to conceive devices potentially more robust to manufacturing imperfections that usually degrade the performance of classical systems and cause decoherence in quantum systems \cite{Klitzing1980}. Topological transport, occurring along boundaries, can be unidirectional and immune to back-scattering, provided that the disorder is not strong enough to close the mobility gap. The unique robustness of topological transport has led to advanced device concepts, including topological delay lines, topological lasers, topological frequency combs, and topological quantum light sources \cite{He2022,Wang2008,Hafezi2011,Mittal2014,St-Jean2017,Bahari2017,Bahari2021,Zhao2019,Smirnova2020}. However, most attention has focused on the treatment of \emph{potential disorder}, i.e., on spatially local random potentials. A different type of disorder, known as \emph{structural disorder}, has long existed in nature, for example in amorphous silicon. In this non-crystalline form of silicon, the local connectivity is preserved, since each silicon atom is bonded to four adjacent neighbors, but long-range structural order is lost. Nevertheless, a mobility gap in amorphous matter has been observed \cite{Weaire1971,Weaire1971a}. More recently, topological phenomena have been observed in non-periodic systems \cite{Agarwala2017,Huang2018,Bandres2016,Zhou2020,Mitchell2018a,Marsal2020}. These findings lead to a natural question to ask: beyond preserving topological properties, can disordered topological systems outperform their periodic counterpart? We propose a structurally disordered system that exhibits a nontrivial topological phase, characterized by a non-uniform synthetic magnetic flux. We show that, in the presence of nonlinearities, the structurally disordered system prevents the ultrafast leakage of energy from topological modes to bulk modes, enhancing nonlinear phenomena. As an example, we demonstrate that the longer confinement of light can lead to  an order of magnitude increase in the generation rate of correlated photon pairs compared to periodic topological platforms.

\section{Results}

\subsection{Disordered linear topological graph}

\begin{figure*}[htbp]
    \centering
    \includegraphics[scale=0.54]{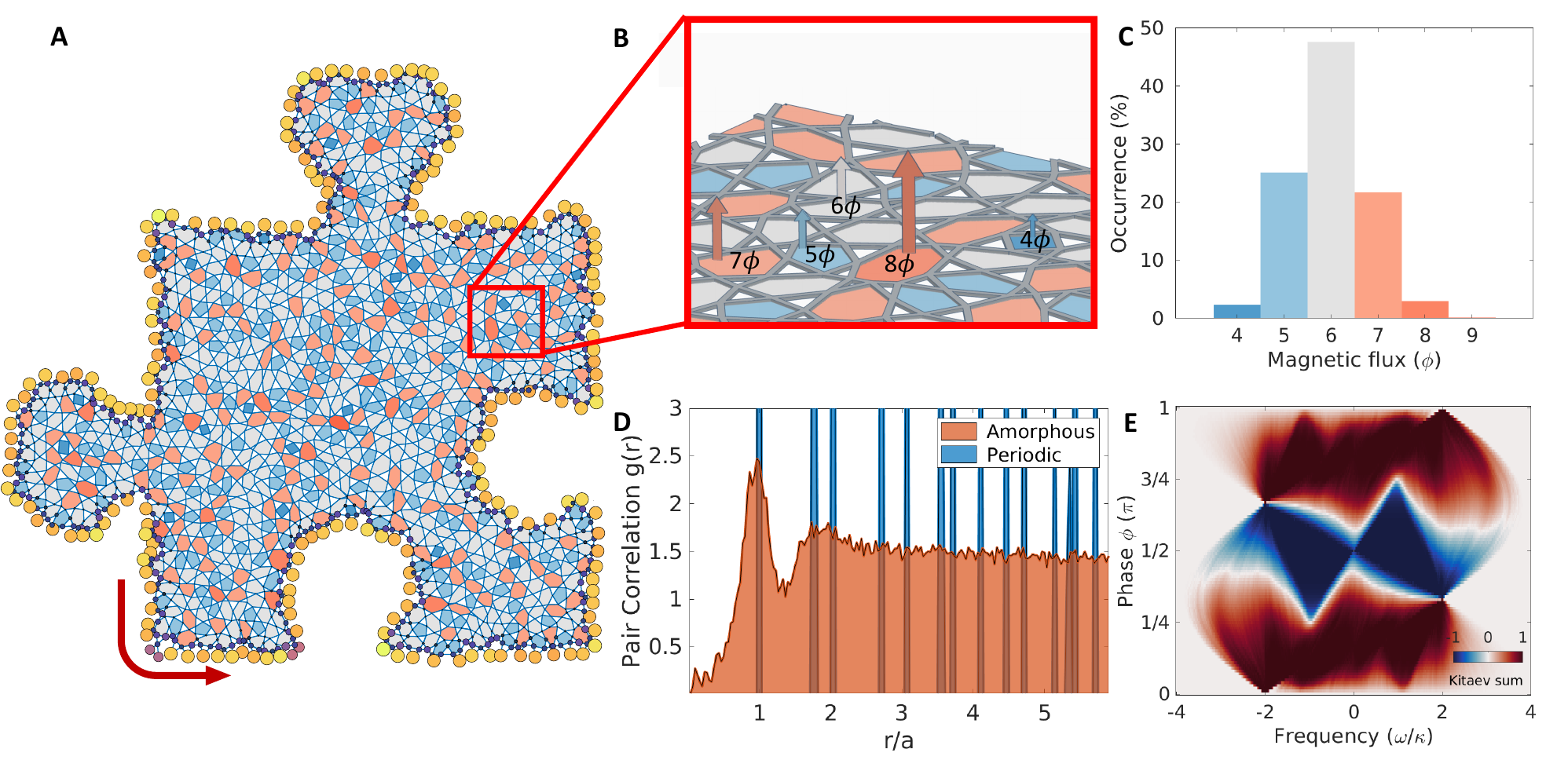}
    \caption{\textbf{Principle and design of amorphous topological graphs.}
    \textbf{a}, Sketch of an amorphous topological graph. The local coordination number $z$ is preserved ($z=4$), while the graph connectivity is different from the periodic counterpart. Different colors indicate polygonal plaquettes with different number of sides.
    \textbf{b}, Zoomed-in view of (\textbf{a}), showing the presence of a non-uniform magnetic field flux. The labels quantify the magnetic flux across each polygonal plaquette, which is equal to the overall hopping phase acquired by a photon through a round-trip around the plaquette.
    \textbf{c}, Distribution of polygonal plaquettes with $N \geq 4$ sides for the graph in (\textbf{a}). The periodic lattices only have hexagons with $N = 6$. 
    \textbf{d}, Pair correlation function $g(r)$ of the amorphous structure in (\textbf{a}), compared to the periodic lattice. The amorphous structure lacks long-range order, as evidenced by the flattened pair correlation function at longer distances $r/a$.
    \textbf{e}, Topological phase diagram for the amorphous structure. The color represents the Kitaev sum calculated over all the modes below different values of the cutoff frequency $\omega/\kappa$, for different hopping phases $\phi$ between adjacent vertices. The result is averaged over $20$ realizations of disorder.}
    \label{fig:fig1}
\end{figure*}

\begin{figure*}[htbp]
    \centering
    \includegraphics[scale=0.64]{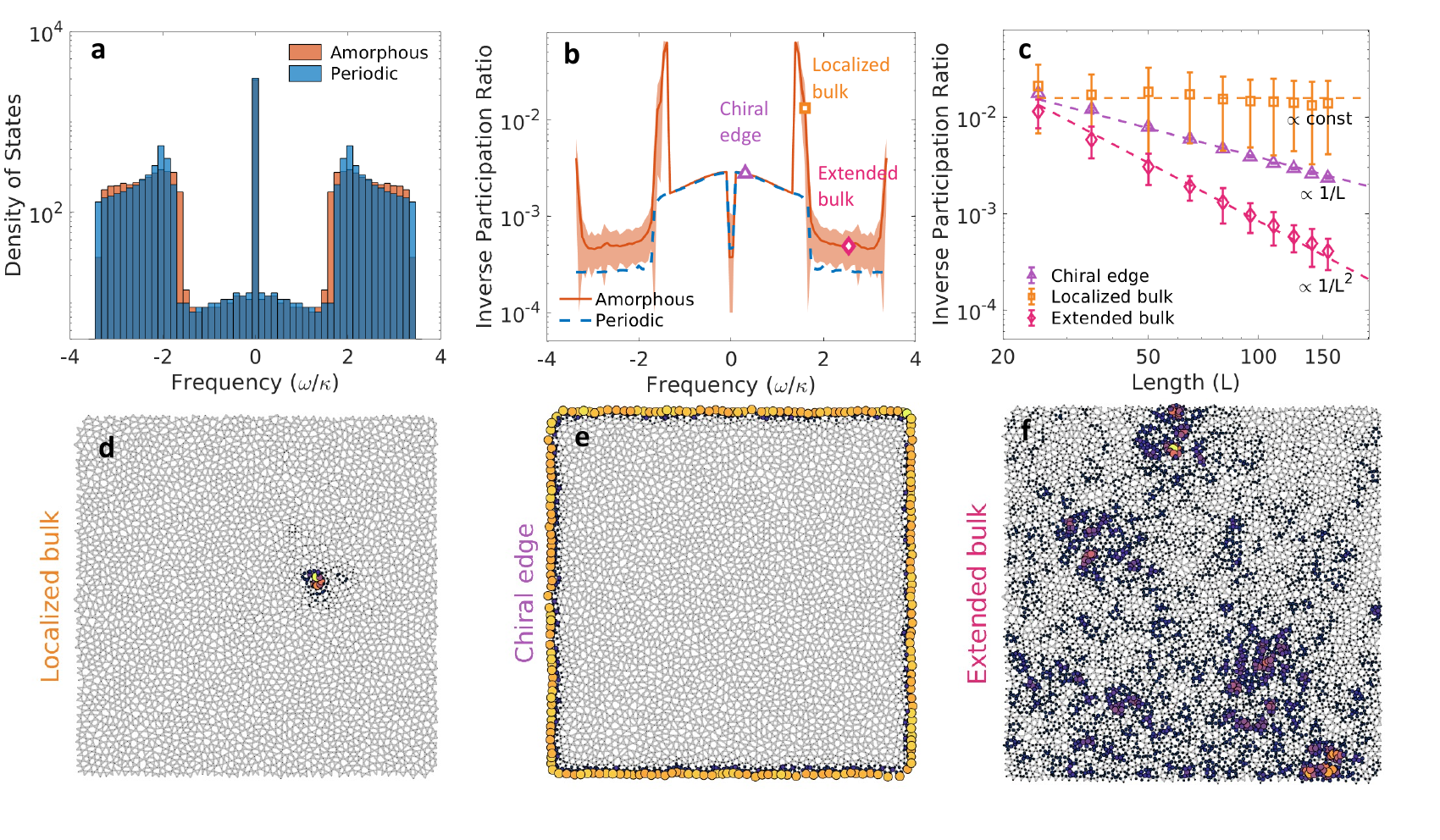}
    \caption{\textbf{Scaling of modes in periodic and amorphous topological graphs.}
    \textbf{a}, Density of states for the amorphous and periodic topological graphs, for a hopping phase $\phi=\pi/2$.
    \textbf{b}, Inverse participation ratio for the eigenmodes of the periodic and amorphous graphs. Three types of modes (chiral edge [CE], localized bulk [LB], and extended bulk [EB]) are observed with distinct inverse participation ratios (IPRs). The shaded area represents the standard deviation for 20 realizations of structural disorder.
    \textbf{c}, Scaling of the IPR with the graph size $L$, for different modes. The results obtained by diagonalizing the linear Hamiltonian in \Cref{eq:linear_hamiltonian} agree well with the theory, and the IPR of modes scales as a constant, $1/L$, and $1/L^{2}$ for LB, CE, and EB modes respectively. The error bars are obtained by averaging over $20$ realizations.
    \textbf{d-f}, Intensity profiles of three representative modes, classified as LB (\textbf{d}), CE (\textbf{e}), and EB (\textbf{f}), showing different localization features. The local light intensity is proportional to the size of the circles at each site, and it is visualized by thermal-like fill color.}
    \label{fig:fig2}
\end{figure*}

The proposed nonlinear amorphous graph, presented in \Cref{fig:fig1}, is constructed by kagomizing a Voronoi diagram obtained from a disk-sampled set of points (Supplementary Information) \cite{Mitchell2018a}. The result is a collage of polygonal plaquettes, each possessing three to nine sides, as sketched in \Cref{fig:fig1}a. Adjacent vertices are then coupled by a directional hopping with uniform magnitude $\kappa$ and phase factor $e^{i\phi}$, shown as the graph edges in the sketch in \Cref{fig:fig1}a. The linear tight-binding Hamiltonian can be written as
\begin{equation}
    \hat{H}_0 = \sum_i \omega_0 \hat{a}_i^\dagger \hat{a}_i -\kappa \sum_{\braket{i,j}} \left( e^{-i\phi} \hat{a}_i^\dagger \hat{a}_j + e^{i\phi} \hat{a}_j^\dagger \hat{a}_i \right) ,
    \label{eq:linear_hamiltonian}
\end{equation}
where $\hat{a}_{i}^{\dagger}$ ($\hat{a}_{i}$) creates (annihilates) a particle on the $i$-th site, $\omega_0$ is the natural on-site frequency, and $\braket{i,j}$ restricts the summation to pairs of nearest-neighbors. The additional hopping phase $\phi$, which can be tuned in a photonic implementation (Supplementary Information), can be interpreted as the Peierls phase resulting from the presence of a synthetic magnetic field. In the model considered here, the synthetic magnetic flux across a polygonal plaquette of the graph depends on the number of edges of the plaquette. Specifically, within each triangular plaquette (white in \Cref{fig:fig1}a), a constant synthetic magnetic field flux of $-3\phi$ is accumulated. In contrast, as sketched in \Cref{fig:fig1}b, the synthetic magnetic field threading each polygonal plaquette with at least four sides is different, and it can vary from $4\phi$ to $9\phi$, proportionally to the number of sides of the polygonal plaquette. Therefore, unlike the anomalous quantum Hall systems which feature a uniform magnetic flux inside hexagonal plaquettes \cite{Hofstadter1976,Haldane1988}, the proposed amorphous system has a non-uniform magnetic flux across different plaquettes based on the real-space connectivity and topology. The statistical distribution of the magnetic flux per plaquette is shown in \Cref{fig:fig1}c, and it can be controlled by changing the filling ratio in the original random disk sampling process (Supplementary Information). By preserving the local connectivity, the generated structure inherently possesses short-range order but lacks long-range order. This can be inferred from \Cref{fig:fig1}d, where a flattened pair correlation function between vertices is observed, unlike periodic structures that exhibit characteristic sharp peaks. Despite the structural disorder, the system shows the hallmarks of nontrivial topology. By controlling the hopping phase, the system undergoes a topological phase transition which opens a nontrivial mobility gap. The topological nature of such a gap can be verified by calculating its Chern number via a topological marker known as ``Kitaev sum'', shown in \Cref{fig:fig1}e \cite{Mitchell2018a,Kitaev2006}, whose value at some frequency $\omega$ expresses the accumulated Chern number from all the bands below the chosen frequency. The complex hopping term $e^{-i\phi}$ allows for a selective tuning of the quantized Kitaev sum across a phase boundary between $-1$ and $+1$, when $\phi$ is set to a value between $0$ and $\pi$. A non-zero topological marker in the gap  implies that a finite-size system will exhibit chiral topological edge states, unidirectionally guided along the physical edge. These states, marked by a low density of states in the topological gap (\Cref{fig:fig2}a), emerge in both periodic and amorphous systems, and are robust to on-site potential disorder as long as the disorder strength is not comparable to the bandgap (Supplementary Information). However, while a strong enough on-site potential disorder will eventually overcome the topological protection of the edge states, an increasing degree of structural disorder will not affect the topological properties of the system, i.e.\@ the edge states are topologically protected irrespective of the degree of structural disorder \cite{Marsal2020}. The regions with a higher density of states in \Cref{fig:fig2}a correspond to bulk bands, including a flat band at zero frequency. We classified the eigenstates $\psi$ by calculating their inverse participation ratio (IPR), defined as $\mathrm{IPR}\left( \psi \right) = \frac{\sum_{i}^{}\left| \psi_{i} \right|^{4}}{\left| \sum_{i}^{}\left| \psi_{i} \right|^{2} \right|^{2}}$, whose scaling law with respect to the lattice size is a measure of the localization of the eigenstates within finite-size systems. The IPR presented in \Cref{fig:fig2}b shows that our amorphous structure features three types of eigenmodes, that are: chiral edge (CE) modes, localized bulk (LB) modes, and extended bulk (EB) modes. The localization and the scaling properties of the modes with the size of amorphous graphs are summarized in \Cref{fig:fig2}c. For a two-dimensional graph with disk sampling domain area $L^2$, the IPR scales as a constant for LB modes, as $1/L$ for CE modes, and as $1/L^2$ for EB modes. Intensity profiles of three representative modes are shown in \Cref{fig:fig2}d-f. The IPRs of the CE modes scale like their periodic counterparts, indicating the existence of topological edge transport channels, while the LB modes, which are a unique feature of the amorphous system, originate from the presence of structural disorder. The LB modes stand out as characteristic peaks in the IPR of \Cref{fig:fig2}b, occurring in the vicinity of the band edges, and they are responsible for the mismatch between the density of states of periodic and amorphous systems (\Cref{fig:fig2}a). The EB modes, that are spatially delocalized, feature low IPRs. The introduction of structural disorder has therefore dramatically changed the localization nature of the bulk modes, introducing the LB modes near the band edges. In addition, the remaining bulk modes are more localized, while the topological nature of the system is preserved.

\subsection{Disordered nonlinear topological graph}

\begin{figure*}[htbp]
    \centering
    \includegraphics[scale=0.54]{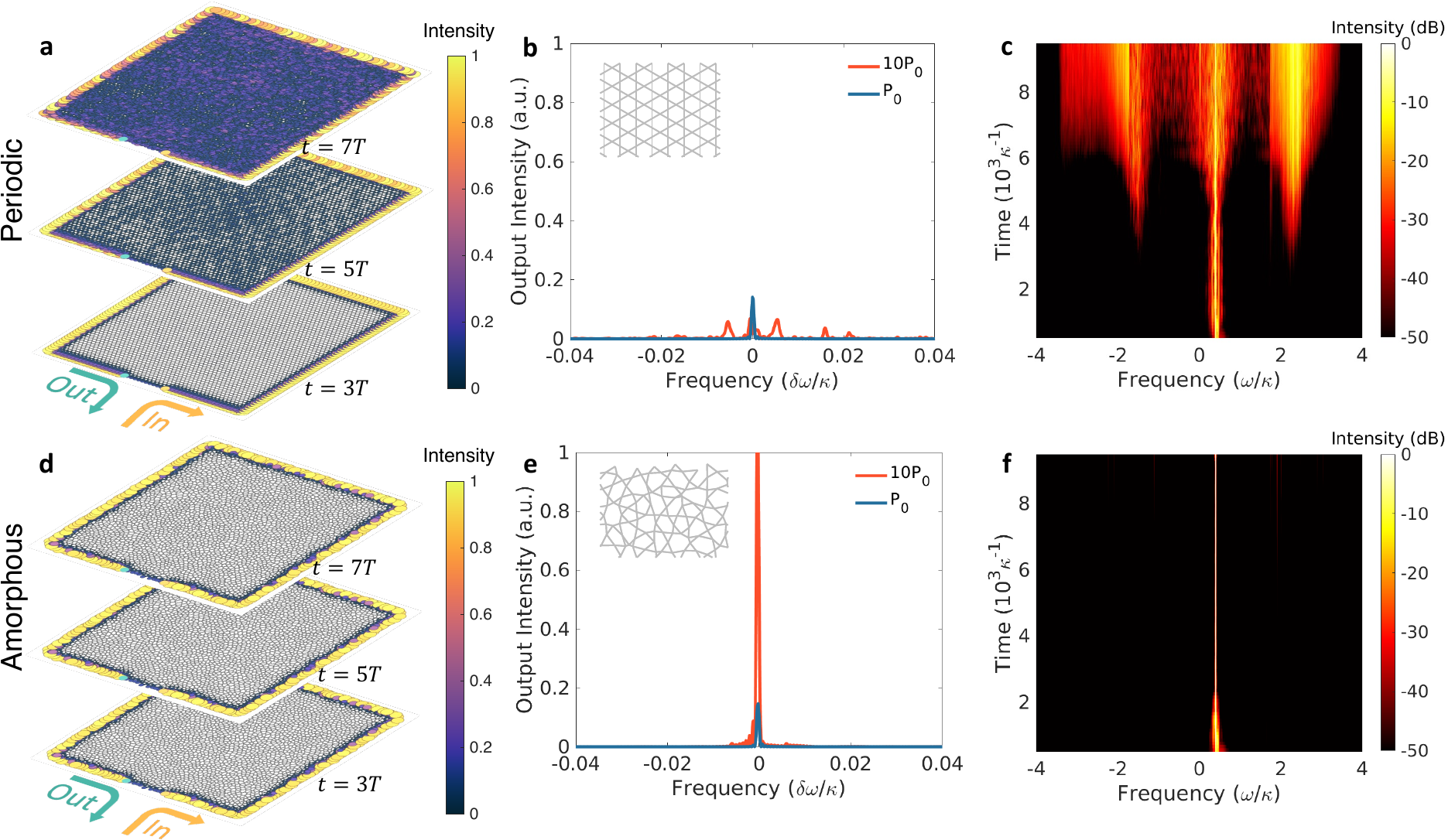}
    \caption{\textbf{Enhanced nonlinear topological transport in amorphous graphs.}
    \textbf{a,d}, Snapshots of the real-space intensity distribution in periodic (\textbf{a}) and amorphous (\textbf{d}) structures, following an initial pulse excitation injected from the input channel, taken at times $3T$, $5T$, $7T$, where $T = 1000\kappa^{-1}$ is approximately the time it takes for the signal to travel from the input to the output port. Energy leaks from the edge to the bulk modes in the periodic graph, while it remains confined along the edge in the amorphous graph for longer times.
    \textbf{b,e}, Power spectrum at the output channel for two different input powers $P_{0}$ and $10P_{0}$, for the periodic (\textbf{b}) and amorphous (\textbf{e}) graphs. The additional peaks in the periodic graph spectrum at $10P_{0}$ pumping correspond to the coupling between adjacent CE modes. The insets show a zoomed-in view of periodic (\textbf{b}) and amorphous (\textbf{e}) graphs.
    \textbf{c,f}, Time-evolution of the power spectrum after injecting an initial signal at an edge mode frequency, obtained via a short-time Fourier transform. In the amorphous case (\textbf{f}), the energy couples to other edge modes or bulk modes at a slower rate compared to the periodic case (\textbf{c}).}
    \label{fig:fig3}
\end{figure*}

\begin{figure*}[htbp]
    \centering
    \includegraphics[scale=0.59]{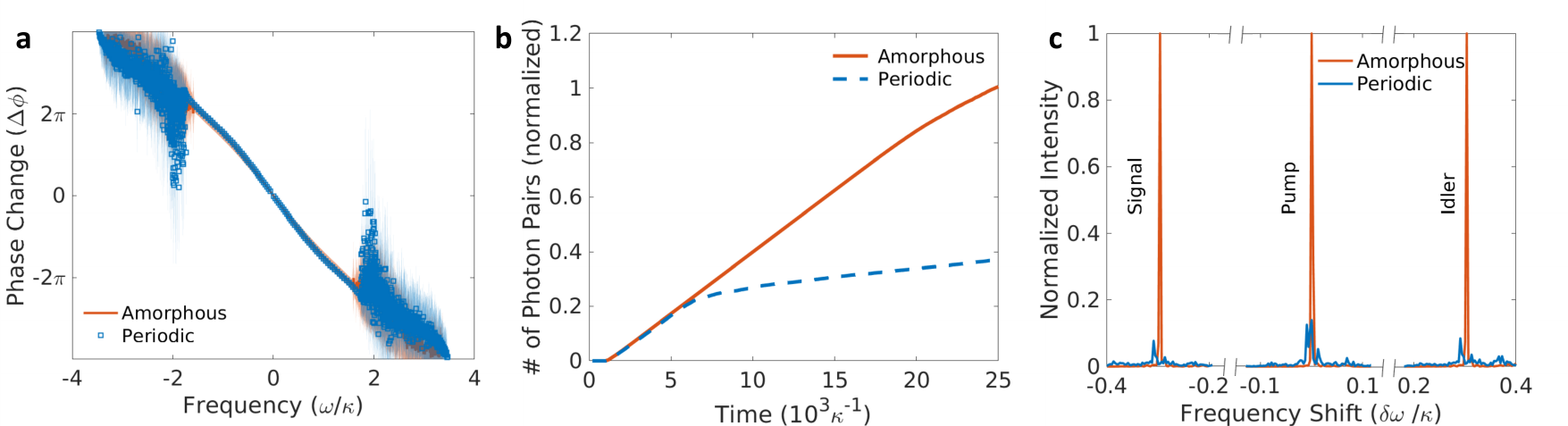}
    \caption{\textbf{Disorder-enhanced topological light generation.}
    \textbf{a}, Effective dispersion relation of amorphous and periodic edge states, obtained by plotting the average phase change between consecutive triangular plaquettes along the edge against the frequency of each eigenmode. For the amorphous system, dispersion broadening is observed due to the aperiodic variation of the phase at different positions along the edge state. Note that, for bulk modes, the quasi-linear phase change breaks down due to the lack of well-defined transport channels.
    \textbf{b}, Time evolution of the total number of generated photon pairs. The photon pair generation rate (slope of the curve) decreases in the periodic system as the signal/idler modes couple to the EB modes, while the generation rate remains high for longer time in the amorphous system.
    \textbf{c}, Example of the normalized spectra of photon pairs generated via a pump at frequency $\omega_p = 0.4\kappa$. The spectra are centered at the pump frequency.}
    \label{fig:fig4}
\end{figure*}

We now consider the nonlinear dynamics of the amorphous topological graph by including multi-particle interactions. As a prototypical example relevant for a wide class of systems, we will focus here on local two-particle interactions, such as the ones that occur between photons in a Kerr medium. The full Hamiltonian $\hat{H}$ describing the nonlinear graph is written by adding a term $\hat{V}$ to the linear Hamiltonian $\hat{H}_{0}$:
\begin{equation}
    \hat{H} = \hat{H}_{0} + \hat{V} ,
    \label{eq:H0_plus_V}
\end{equation}
where
\begin{equation}
    \hat{V} = U_0\sum_i \hat{n}_i^2,
    \qquad
    \hat{n}_i = \hat{a}_i^\dagger \hat{a}_i\,.
    \label{eq:Kerr_nonlinearity}
\end{equation}
In \Cref{eq:Kerr_nonlinearity}, $U_{0}$ is a material-dependent strength of the nonlinearity and $\hat{n}_{i}$ is the particle number operator at site $i$. The time dynamics of the nonlinear system is obtained by explicitly integrating the time-evolution equations. The periodic and the amorphous topological systems are both driven with the same amplitude, which is spectrally peaked within the topological bandgap. Snapshots of the intensity distribution at different times $t$ are shown for the periodic (\Cref{fig:fig3}a) and amorphous (\Cref{fig:fig3}d) graphs. The energy of the propagating excitation is confined near the boundary at early time stages in both cases ($t_{0} < 3T$), where $T = 1000\kappa^{- 1}$ is approximately the time the signal takes to travel from the input to the output port. In the periodic case (\Cref{fig:fig3}a), the excited CE modes couple to other CE modes, and then leak towards the bulk as the EB modes are fed ($t = 5T$). This contrasts with the amorphous case, where energy is confined within the CE modes at $t = 5T$, as shown in \Cref{fig:fig3}d. The difference in energy transport in the presence of nonlinearity is quantified by probing the transmission at the output port for different injected power levels, as shown in \Cref{fig:fig3}b,e. For a relatively weak pumping power $P_{0}$, the transmission spectrum is similar in the periodic and amorphous systems. As the pumping power increases, the edge transport channel breaks down in the periodic system due to the nonlinearity-induced coupling, while the transmission in the amorphous system maintains chiral edge propagation, leading to an almost tenfold increase in the peak power difference at the output. The stronger coupling between adjacent CE modes in the periodic case is confirmed by the presence of additional side peaks in the transmission spectrum. The evolution of the two systems in the time-frequency domain, obtained via a short-time Fourier transform, is presented in \Cref{fig:fig3}c,f. In the periodic system, at early times, the initialized CE mode couples to the spectrally closest CE modes, resulting in frequency broadening. As time evolves, the coupling between the excited CE modes and EB modes, induced by the presence of the nonlinearity, ignites bulk modes, which in turn excite other EB and CE modes. The result, at long times, is a dramatically broadened spectrum. The amorphous system in \Cref{fig:fig3}f, however, shows a strikingly different behavior, displaying both reduced oscillations between adjacent CE modes as well as a notably suppressed appearance of additional modes in the spectrum, achieving an almost unperturbed propagation for much longer times.

\section{Discussion}

The increased isolation that the injected CE mode experiences in the amorphous system can be understood as an interplay of different mechanisms. First, the broken periodicity resulting from the introduction of structural disorder precludes us from identifying a well-defined momentum for the EB modes, hampering the fulfilment of phase-matching conditions for the nonlinearity-induced coupling, and suppressing the initial oscillations between CE modes. Second, according to Fermi’s golden rule, the initially excited bulk modes will be located around the peak of density of states in \Cref{fig:fig2}a, close to the band edges. These modes have an EB nature in the periodic system, but, with the introduction of structural disorder, some of them become LB modes in the amorphous system \cite{Mott1967}. The localization of LB modes, then, delays the nonlinearity-induced propagation from CE to EB modes, with the latter being excited at much later times. The eventual propagation of the signal to the EB modes is in fact delayed by the introduction of structural disorder, but not completely suppressed, as an inevitable consequence of the presence of the nonlinearity.

The spatial and spectral energy confinement in the presence of nonlinearity can be used, for example, to enhance the efficiency of quantum topological photon-pair generation via spontaneous four-wave mixing (sFWM) in optical systems \cite{Mittal2018a,Mittal2021}. In this non-linear four-photon process, a pump signal is injected into the system via an input port, and the correlated photon pairs generated within the topological bandgap are guided towards the output port along the boundaries. The positions of input and output ports are chosen as shown in \Cref{fig:fig3}a,d to increase the distance travelled by the pump signal, thereby maximizing the photon pair generation efficiency. The system is described by the following Hamiltonian \cite{Mittal2018a,Ong2013b,Kumar2014},
\begin{equation}
    H_{SI} = H_0^{(S)} +H_0^{(I)} + \sum_i \chi_i {\hat{a}^{(S)}_i}^\dagger {\hat{a}^{(I)}_i}^\dagger
    \label{eq:H_fwm}
\end{equation}
where the last term creates correlated pairs of photons, called signal (S) and idler (I), at a site-dependent rate $\chi_{i} = \chi_{0}\hat{a}_{i}^{(P)}\hat{a}_{i}^{(P)}$ which depends both on the pump (P) strength and on the optical nonlinearity $\chi_{0}$.  The terms $H_{0}^{\left( S/I \right)}$ are the bare Hamiltonians of the signal and idler photons, respectively. The effectiveness of this process typically relies on working with a quasi-linear dispersion to satisfy both energy and momentum conservation. As shown in \Cref{fig:fig4}a, the topological dispersion in our system fulfills this requirement. The photon pair generation efficiency is presented in \Cref{fig:fig4}b. Both the periodic and the amorphous system generate a similar number of photon pairs at early times. At later times, however, the efficiency of the periodic system drops due to the increased coupling between CE and EB modes, while the amorphous system remains efficient. The enhanced spectra of signal, pump and idler are presented in \Cref{fig:fig4}c. In the periodic case, the self-modulation of the pump leads to a reduced lifetime of the edge modes at the excitation frequencies, hence the photon pair generation rate is also reduced, while in the amorphous case, the pumped edge states can generate more photon pairs.

We proposed and demonstrated an amorphous topological platform (graph) to enhance non-linear phenomena. The disordered topological graphs are based on the control of a synthetic magnetic field threading through different polygonal plaquettes and leading to a non-uniform flux that gives rise to the non-trivial topology of the graph. The non-linear responses of such amorphous topological graphs outperform their periodic counterparts owing to the emergence of localized bulk modes and to a reduced phase matching, enabling the ultrafast guiding of energy, protected against leakage. The accumulation of energy in the topologically protected channel enhances the generation of photon pairs. The proposed scheme for enhancing non-linear responses via the introduction of structural disorder in topological systems will have a broad range of applications, including novel devices for robust light-based information processing and computing.

\section*{Materials and Methods}

\subsection*{Generation of the amorphous topological graph}

The amorphous graphs are generated via a three-step procedure to maintain a fixed coordination number while implementing the structural disorder. First, disk sampling with a random seed is used to introduce disorder in the system. Then, the centers of the disks are used to generate a Voronoi graph with a fixed coordination number of $3$. The amorphous Kagome graph is finally constructed by connecting the centers of adjacent edges around each vertex of the Voronoi diagram. The generated Kagome graph has a coordination number of 4, which is the same as that of a periodic Kagome lattice. In the disk sampling process, the filling ratio can be used to control the degree of disorder. In the amorphous structures simulated in this work, a filling ratio of $\eta=0.45$ is used with a domain size of $L/r=250$ in the disk sampling step.

\subsection*{Calculation of the topological index}

The topology of the amorphous graph is explored via the Kitaev sum. The graph is first triparted into $3$ adjacent spatial regions $A$, $B$, $C$ that share a common vertex at the center of the graph and that cover $1/4$ of its area. Then, the Kitaev sum at frequency $\omega_c$ is calculated as
\begin{equation}
    \nu(P) = 12\pi\sum_{i\in A}\sum_{j\in B}\sum_{k\in C} \Big( P_{ij}P_{jk}P_{ki} - P_{ik}P_{kj}P_{ji} \Big) \, ,
\end{equation}
where $P$ is the projection operator onto the eigenmodes below $\omega_c$ and $i$, $j$, $k$ are site indices.

\subsection*{Time-domain simulations}

The time-domain simulations were performed on MATLAB\textsuperscript{®} via explicit integration of the semiclassical evolution of the particle fields, using a 4th order Runge-Kutta scheme in order to ensure numerical stability.
The four-wave mixing process was simulated in the undepleted pump approximation, under which the fields evolve via a time-dependent Hamiltonian that can be represented as the following matrix \cite{Ong2013b}:
\begin{equation}
    H_{SI}(t) = 
    \begin{bmatrix}
        H_0^{(S)} & C(t)\\
        C(t)^\dagger & H_0^{(I)\dagger}
    \end{bmatrix},
\end{equation}
where $C_{ij} = \chi_{i}(t)\delta_{ij}$ and the state vector is expressed as $\psi=\left[a_1^{(S)},\ldots,a_N^{(S)},a_1^{(I)\dagger},\ldots,a_N^{(I)\dagger}\right]^\mathsf{T}$. The outputs are Fourier-transformed to obtain the spectral information.

\section*{Acknowledgments}

\textbf{Funding:} This work was supported by the Moore Inventor Fellows programme, the Bakar Fellowship at UC Berkeley, the NSF QLCI programme through Grant No.\@ OMA-2016245, the NSF QuIC-TAQS award 2137645. It was partially supported by the Office of Naval Research (ONR) grant N00014-20-1-2723, the ONR grant N00014-22-1-2651, and the US Army Research Office (ARO) grant W911NF2310027. A.A.\@ and J.M.\@ were supported by NSF DMR-1918065.  E.D.\@ was supported by an NSF Graduate Fellowship. J.M.\@ acknowledges additional support from a Simons Investigatorship. \textbf{Author contributions:} Z.J.\@ and B.K.\@ conceived the project. Z.J.\@ and M.S.\@ performed the theoretical calculations. M.S.\@, Z.J.\@, and B.K.\@ wrote the manuscript with inputs from A.A.\@, W.R.\@, E.D.\@, and J.M. All authors analyzed the results and contributed to discussions. B.K.\@ and J.M.\@ supervised the project. \textbf{Competing interests:} The authors declare no competing interests. \textbf{Data and materials availability:} All data needed to evaluate the conclusions in the paper are present in the paper and/or the Supplementary Materials.

%%%%%%%%%%%%%%%%%%%%%%%%%%%%%%%%%%%%%%%%%%%%%%%%%%%%%%%%%%%
% References                                              %
%%%%%%%%%%%%%%%%%%%%%%%%%%%%%%%%%%%%%%%%%%%%%%%%%%%%%%%%%%%

% Extra references from Supplementary Material
\nocite{Ohgushi2000}

%\clearpage                              % Avoids references between figures
\vbox{}                                 % Avoids overlap of references symbol

\bibliographystyle{naturemagwithdoi}    % Default: apsrev4-1 | See https://tex.stackexchange.com/q/76028
\typeout{}                              % See Overleaf's help page
\bibliography{references}

% Include pre-compiled Supplementary Material
\includesupplementary

\end{document}

\title{Supplementary Information for\texorpdfstring{\\}{} ``Disordered topological graphs enhancing nonlinear phenomena''}

\author{Zhetao Jia}
\affiliation{Department of Electrical Engineering and Computer Sciences, University of California, Berkeley, CA 94720, USA}

\author{Matteo Seclì}
\affiliation{Department of Electrical Engineering and Computer Sciences, University of California, Berkeley, CA 94720, USA}

\author{Alexander Avdoshkin}
\affiliation{Department of Physics, University of California, Berkeley, California 94720, USA}

\author{Walid Redjem}
\affiliation{Department of Electrical Engineering and Computer Sciences, University of California, Berkeley, CA 94720, USA}

\author{Elizabeth Dresselhaus}
\affiliation{Department of Physics, University of California, Berkeley, California 94720, USA}

\author{Joel Moore}
\affiliation{Department of Physics, University of California, Berkeley, California 94720, USA}
\affiliation{Materials Sciences Division, Lawrence Berkeley National Laboratory, 1 Cyclotron Road, Berkeley, CA 94720, USA}

\author{Boubacar Kanté}
\email[]{bkante@berkeley.edu}
\affiliation{Department of Electrical Engineering and Computer Sciences, University of California, Berkeley, CA 94720, USA}
\affiliation{Materials Sciences Division, Lawrence Berkeley National Laboratory, 1 Cyclotron Road, Berkeley, CA 94720, USA}

%\date{May 11, 2023}

\maketitle

\section{Periodic kagome lattice and band structure}

\begin{figure*}[ht]%[htbp]
    \centering
    \includegraphics[scale=0.18]{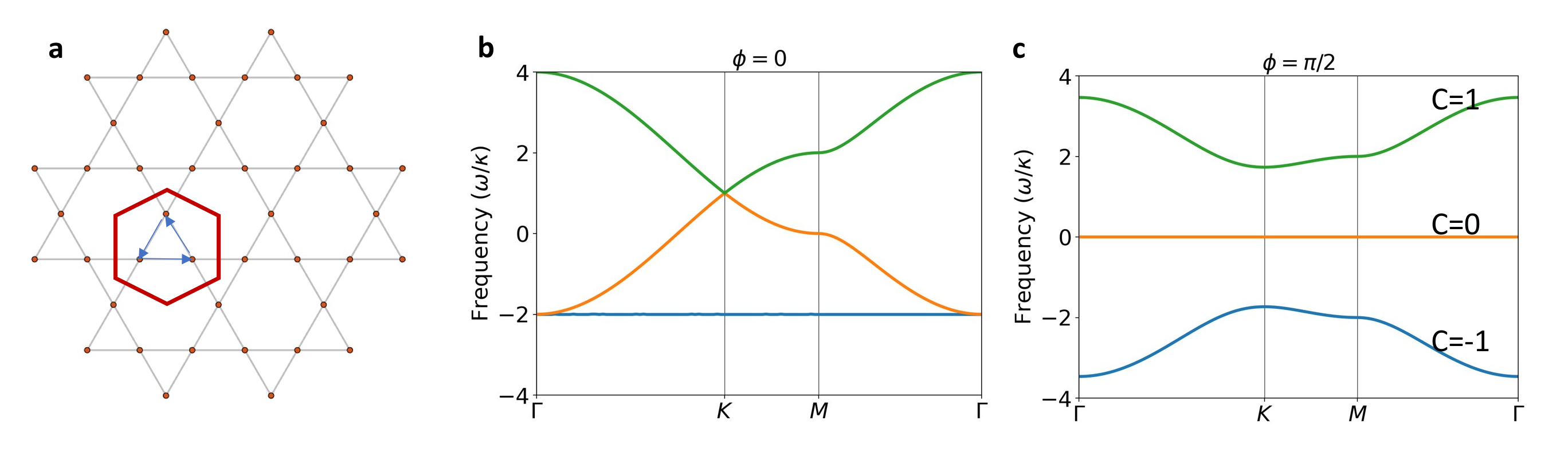}
    \caption{\textbf{Topological kagome lattice.}
    \textbf{a}, Structure of a finite-size kagome lattice with a directional hopping phase. The unit-cell, containing three sites, is highlighted by the red hexagon.
    \textbf{b}, Band diagram of a kagome lattice
    with a null directional hopping phase ($\phi = 0$), where the coupling between sites is real. 
    \textbf{c}, Band diagram of a topological kagome lattice with $\phi = \pi/2$, which corresponds to a complex hopping $e^{-i\phi}=-i$. Topological bands with nonzero Chern numbers appear.}
    \label{fig:figS1}
\end{figure*}

To investigate the proposed amorphous topological graph, we first study its periodic counterpart.\cite{man:Ohgushi2000} In \Cref{fig:figS1}, a finite-size periodic kagome lattice is shown. Its unit cell is highlighted within the red hexagon, with arrows indicating a directional hopping phase $e^{-i\phi}$. The Hamiltonian of the system is of the same form as Eq.~1 in the main text, but the connectivity is different in periodic and amorphous systems. The difference is encoded in the positions of nonzero matrix elements in the Hamiltonian of the tight-binding model, and cannot be eliminated by a permutation of the basis elements, i.e. the two graphs are non-isomorphic. In the periodic system, the hopping phase $\phi$ determines the band structure, and opens a topological gap when the hopping phase is tuned from $\phi = 0$ to $\phi = \pi/2$. The nontrivial topological properties can be illustrated by the Chern number, calculated by integrating the Berry curvature in momentum space. The calculated topological indices of three bands are labeled in \Cref{fig:figS1}c, with $-1, 0, 1$ from lower to higher bands, respectively.

\section{Generation of the topological amorphous graph}
\label{supp:sec:generation}

\begin{figure*}[ht]%[htbp]
    \centering
    \includegraphics[scale=0.2]{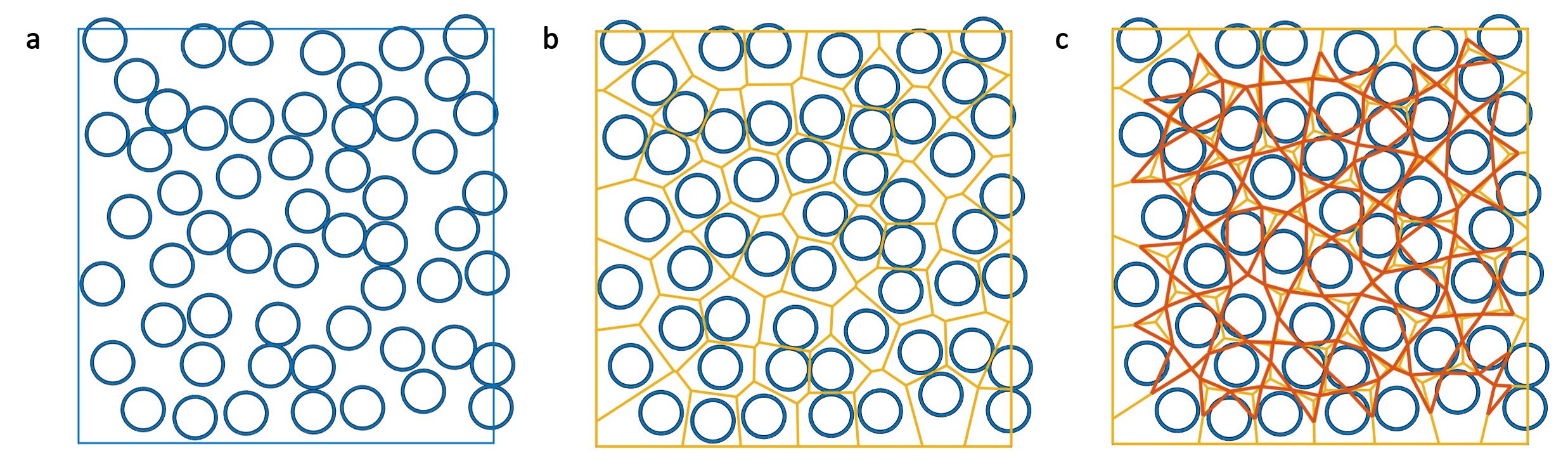}
    \caption{\textbf{Amorphous graph generation procedure.}
    \textbf{a}, Disks are randomly sampled within the domain with periodic boundary conditions. Each disk has a fixed radius and does not overlap with its neighbors.
    \textbf{b}, Voronoi diagram is generated based on the disk centers.
    \textbf{c}, The centers of edges sharing to the same vertices are connected to generate a kagome-like graph.}
    \label{fig:figS2}
\end{figure*}

The topological amorphous graph is generated by tessellation and local triangularization of randomly sampled points in two-dimensional space.\cite{man:Mitchell2018a} The initial random points can be generated, for example, with disk sampling (\Cref{fig:figS2}a) in a domain with area $L^2$ and periodic boundary conditions. The filling ratio $\eta$, defined as the ratio of area covered by disks over the area of the sampling domain, is the parameter used to control the strength of structural disorder. The Voronoi diagram is used to partition space, and create a graph with a fixed coordination number $N=3$ (\Cref{fig:figS2}b). The amorphous graph is created by a kagomization process, in which the centers of the edges sharing the same vertices are connected. The resulting graph has a local coordination number $(N=4)$ in \Cref{fig:figS2}c. Because of the coordination number $N=3$ for the Voronoi diagram, the generated amorphous graph consists of local triangles around the vertices of the Voronoi diagram, but the tessellated regions consist of polygons with different number of sides. The periodic lattice in \Cref{fig:figS1}a can also be generated by the same procedure, with the difference that, instead of starting from a random point set, we use a triangular lattice. The triangular lattice arrangement, indeed, is the most compact configuration achievable via disk sampling, characterized by a maximum filling ratio $\eta_{max} = 0.9069$. The amorphous graphs are generated with a fixed filling ratio $\eta = 0.45$ for the disk sampling process. To make sure the sampling can be achieved within reasonable simulation time, the amorphous graph used in Fig. 3 and 4 has a  domain size $L/r=250$, where $r$ is the radius of the disk. To compare with the periodic system, the total number of sites is controlled. Due to the randomness of the disk sampling process, the exact number of sites may differ, but the difference is less than $1\%$ among realizations. 

\begin{figure*}[h]%[htbp]
    \centering
    \includegraphics[scale=0.2]{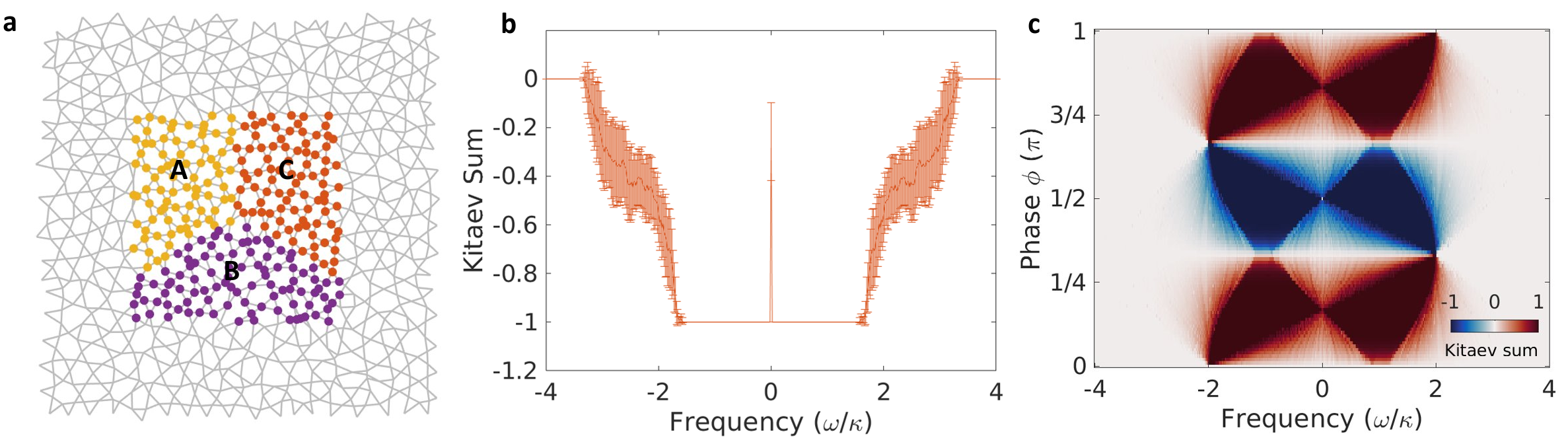}
    \caption{\textbf{Kitaev sum calculation.} \textbf{a}, Calculation of the Kitaev sum in the amorphous structure. To calculate the local topological index, the integration region is subdivided into three sections, marked here with different colors. \textbf{b}, Kitaev sum averaged over 20 random realizations of a graph with a hopping phase $\phi = \pi/2$. \textbf{c}, Kitaev sum calculated on a finite-size periodic lattice.}
    \label{fig:figS3}
\end{figure*}

\section{Chern number calculation in the amorphous topological graph}
\label{supp:sec:chern}

The topology of the graph is probed by the Kitaev sum, a topological index defined as \cite{man:Kitaev2006,man:Mitchell2018a}
\begin{equation}
    \nu(P) = 12\pi\sum_{i\in A}\sum_{j\in B}\sum_{k\in C} \Big( P_{ij}P_{jk}P_{ki} - P_{ik}P_{kj}P_{ji} \Big) \, ,
\end{equation}
where $P$ is the projection operator onto the eigenmodes below the cut-off energy, $A$, $B$, $C$ are three spatial regions shown in \Cref{fig:figS3} and $i$, $j$, $k$ are sites within corresponding regions. 

The summation region is fixed to span half of the side length of the lattice, as shown by the square in \Cref{fig:figS3}, and it is divided into three regions $A, B,$ and $C$ represented by different colors. To study the phase transition under different hopping phases $e^{-i\phi}$, the cut-off frequency is swept, and the topological index is averaged over $20$ random realizations of the amorphous graph. The phase diagram in Fig.~2 of the main text shows that a Kitaev sum of $\pm 1$ can be reached by tuning the hopping phase. As an example, the standard deviation of the Kitaev sum with phase $\phi = \pi/2$ is shown in \Cref{fig:figS3}b, revealing that the Kitaev sum remains $-1$ within the topological bandgap with negligible variance across realizations of structural disorder. As a comparison, the Kitaev sum of a similar periodic system is show in \Cref{fig:figS3}c. The periodic phase diagram has sharper boundaries due to lack of the localized bulk modes near the band edges. The bandgap closes for phase values of $\phi = 0, 1/3\pi, 2/3\pi$ and $\pi$, in agreement with \Cref{fig:figS1}b.

\section{Robustness to on-site disorder}
\label{supp:sec:onsite_disorder}

\begin{figure*}[hb]%[htbp]
    \centering
    \includegraphics[scale=0.23]{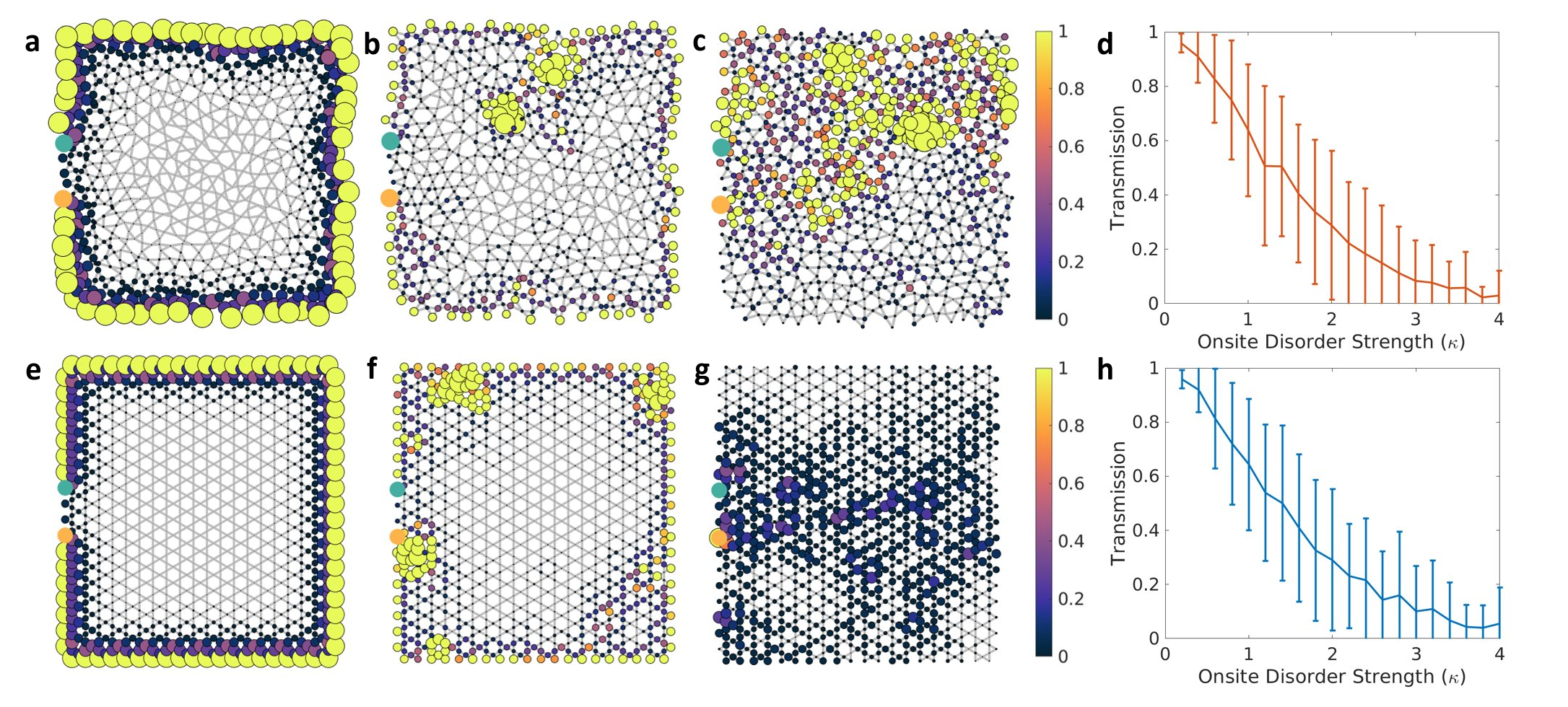}\\
    \caption{\textbf{Robustness to on-site disorder.} \textbf{a--c}, Topological transport on an amorphous graph with different on-site potential disorder. The disorder strength is characterized by the standard deviation of a Gaussian distribution with $\sigma/\kappa = 0$ (\textbf{a}), $0.4$ (\textbf{b}), $1.6$ (\textbf{c}). The transmission with increasing on-site potential is summarized in \textbf{d} by averaging over $100$ realizations of disorder. \textbf{e--h}, Topological transport on a periodic graph with potential disorder, similar to \textbf{a--c}.
    }
    \label{fig:figS4}
\end{figure*}

Despite the structural disorder, the amorphous system is robust to on-site (potential) disorder, up to a certain strength. In actual physical systems, such disorder may stem from fabrication imperfections that e.g.\@ shift the natural frequency of the resonators employed as lattice sites. The on-site disorder is assumed here to follow a Gaussian distribution, characterized by the standard deviation $\sigma$ which we sweep for the numerical simulations. The existence of a transport channel along the topological edge state is probed at the output site, with a single frequency excitation on the input port. The transport can be visualized by the intensity distribution at the steady state, shown in \Cref{fig:figS4}a-c. As the strength of on-site disorder increases, the edge state couples to localized modes near the boundary and the transport efficiency is reduced to half when the standard deviation of the on-site disorder is equal to the coupling strength between sites, $\kappa$. As shown in \Cref{fig:figS4}d, h, the transmission is further reduced to less than $5\%$ as the disorder strength reaches $3\kappa$. 

\section{Photonic implementation parameters}
\label{supp:sec:implementation}

We envision a potential implementation of an amorphous topological graph on a silicon nitride photonic platform.\cite{man:Mittal2018a} The photon pair generation efficiency and the corresponding timescale are then expressed in relation to the coupling strength between adjacent sites. In practice, we assume a coupling strength of $\kappa=\SI{40}{\GHz}$, which translates to a timescale of $\kappa^{-1} = \SI{0.025}{\ns}$. For the typical graph sizes used in our simulation, the time it takes for the light to travel from the input to the output port is roughly $T = 1000\kappa^{-1} = \SI{25}{\ns}$. For a typical single-mode silicon nitride waveguide, the effective area is of the order of $A = \SI{0.5}{\um} \cdot \SI{0.2}{\um} = \SI{e-13}{\square\meter}$. For a pump power $\SI{1}{\mW}$, the intensity inside the waveguide is $I=\SI{e10}{\W\per\square\meter}$. With a Kerr coefficient of silicon nitride around $n_2 =\SI{e-18}{\per\W\square\meter}$, the nonlinear effect can be quantified by the relative change of refractive index as $\delta n/n=10^{-5}$ within a ring resonator with quality factor of order $10^3$. The corresponding frequency shift of the resonance of $n_2 I/\kappa = 0.1$ is used in the simulations with the normalized nonlinear coefficient $U/\kappa = 0.03$ and the coupling coefficient of the input coupler as $\kappa_{ex} = \kappa^2/2$.

\section{Time-domain four-wave mixing simulation}
\label{supp:sec:FWM}

\begin{figure*}[h]%[htbp]
    \centering
    \includegraphics[scale=0.4]{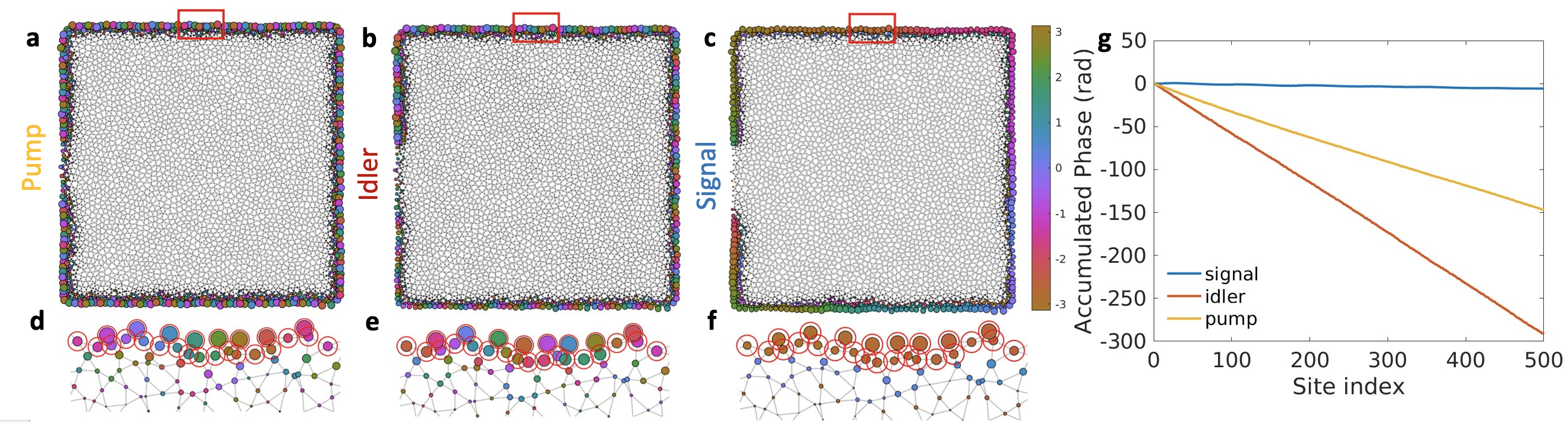}
    \caption{\textbf{Four-wave mixing field profiles.} The light is injected into the input port and coupled out at the output port, similar to Fig.~3 of the main text. \textbf{a--c}, Field profiles for pump (\textbf{a}), idler (\textbf{b}), and signal (\textbf{c}), where the color (size) represents the phase (amplituede) at each site. \textbf{d--f}, Zoomed-in view of the field distribution at sites along the edge. Red circles indicate the sites used for the phase calculation in \textbf{g}. \textbf{g}, Accumulated phases over 500 edge sites for pump, idler and signal fields.}
    \label{fig:figS5}
\end{figure*}

In the spontaneous four-wave mixing process, signal and idler photon pairs are generated and transported along the chiral edge states. In practice, e.g.\@ in a microrings platform, the generated photon pairs can be hosted at different resonance frequencies of the ring resonators.\cite{man:Mittal2018a} In this case, the frequencies of the generated light differ from the pump frequency by a free spectral range, so that the pump light can be filtered out. The Hamiltonian of signal and idler frequencies in the linear regime is assumed to be the same as the pump, which have chiral transport channels along the edge, shown in Fig.~2 of the main text. The four-wave mixing Hamiltonian can be represented as\cite{man:Ong2013b}
\begin{equation}
    H_{SI} = 
    \begin{bmatrix}
        H_S & C\\
        C^\dagger & H_I^\dagger
    \end{bmatrix},
\end{equation}
where $H_S=H_I=H_0$ and $C_{ij} = \chi_{i}(t)\delta_{ij}$. Here, the four-photon interaction strength at site $i$, represented as $\chi_i(t)$, is time dependent, and proportional to the intensity of the pump at site $i$. The linear dynamics of signal and idler frequencies are governed by the diagonal part of the Hamiltonian, while the pump-induced photon pair generation is represented as the off-diagonal component. The four-wave mixing process is simulated in the time domain under the undepleted pump assumption, as the pump experiences self-modulation, but it is not affected by signal or idler photons due to their relatively low intensity. To quantify the photon pair generation efficiency, a small signal at idler frequency is injected into the system, and the signal at the output port is recorded and Fourier-transformed to obtain the spectral information. The evolution of the generated signal photon spectrum over time is calculated by a short-time Fourier transform of the output signal. A snapshot of the field distribution in the amorphous system is shown in \Cref{fig:figS5}, where the pump, idler and signal field profiles are plotted in panels a, b, c, respectively. In the photon pair generation process, the momentum and energy conservation has to be satisfied. In the amorphous system, the effective momenta of the excited edge modes can be estimated by calculating the phase differences between adjacent sites along the edge, which are visualized as zoomed-in views in \Cref{fig:figS5}d--f.  The accumulated phase differences are plotted against the site index in \Cref{fig:figS5}g, from which we can extract the average phase shift per site as $\Delta\phi_p = -0.293, \Delta\phi_i = -0.583, \Delta\phi_s = -0.012$. The momentum conservation can then  be verified in terms of the phase differences as $2\Delta\phi_p = \Delta\phi_s + \Delta\phi_i$, which is a natural consequence of the four-wave mixing simulation. 

\section{Different amorphous realizations and truncations}
\label{supp:sec:differentrealizations}

In the main text, we show the generated amorphous topological graph, following the generation procedure described in \Cref{supp:sec:generation} with a square boundary shape. Here, we show that the enhanced transport in amorphous system is independent of the specific realization of structural disorder, and is not limited to a square boundary truncation. As an example, we implement different realizations of the amorphous graphs, and compare them with the periodic systems truncated at different orientations. For each system, the same signal is injected into the structure and the short-time Fourier transform is computed, in order to capture the frequency information of the lattice intensity over time. For the amorphous graph, the short-time Fourier transform is similar to the square cut in Fig.~3 of the main text, as different truncations of the graphs have negligible effect on the amorphous system. The periodic system, however, shows different response under different truncation orientations. Nevertheless, the energy confined in the edge state of the amorphous system decays at a much slower rate compared to the periodic system, independently of the truncation choice.

\begin{figure*}[ht]%[htbp]
    \centering
    \includegraphics[scale=0.25]{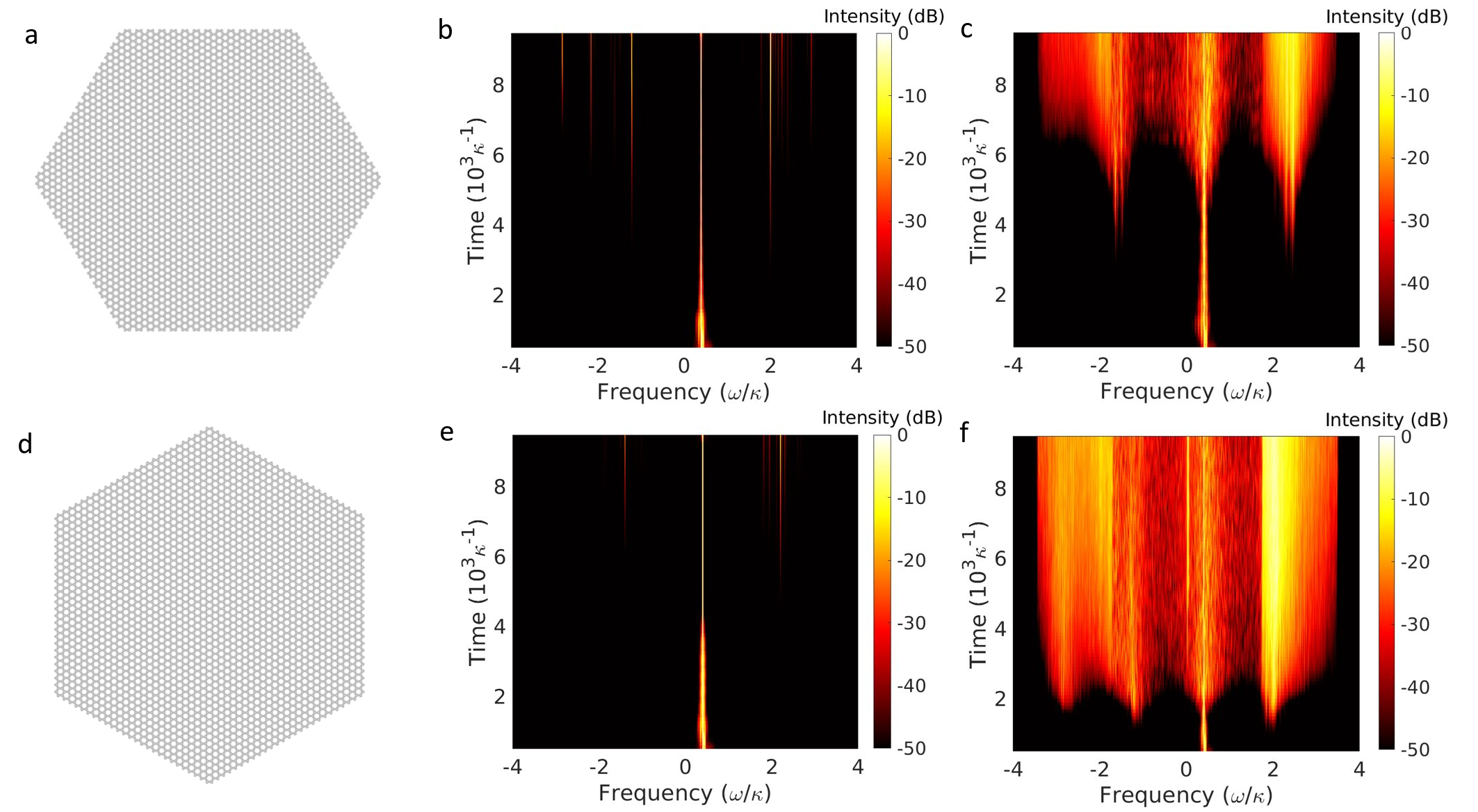}
    \caption{\textbf{Different realizations and truncations.} The periodic system is truncated along different lattice orientation directions, and the short-time Fourier transform of the signal is compared to the amorphous system. \textbf{a}, Armchair-like truncation of the periodic lattice. \textbf{b}, Short-time Fourier transform of energy transported in an amorphous system with nonlinearity. \textbf{c}, Short-time Fourier transform for transport in the periodic system. \textbf{d--f}, Same as \textbf{a--c}, but with a zig-zag-like truncation.}
    \label{fig:figS6}
\end{figure*}

%% file: extra/includesupplementary.tex
\usepackage{ifxetex}
\usepackage{pdfpages}
\ifxetex\else\usepackage{pax}\fi
\usepackage{forloop}
\makeatletter\AtBeginDocument{\let\LS@rot\@undefined}\makeatother
\ifxetex\newcount\pdflastximagepages\def\pdfximage#1{\pdflastximagepages=\XeTeXpdfpagecount"#1"\relax}\fi
\pdfximage{\supplementfilename.pdf}
\def\numbersupplementpages{\the\pdflastximagepages}
\newcommand{\includesupplementary}{
    \onecolumngrid
    \newpage
    \pdfbookmark[0]{Supplementary Information}{suppl}
    \newcounter{x}
    \forloop{x}{1}{\value{x} < \numbersupplementpages \OR \equal{\value{x}}{\numbersupplementpages}}
    {
        \includepdf[pages=\arabic{x}]{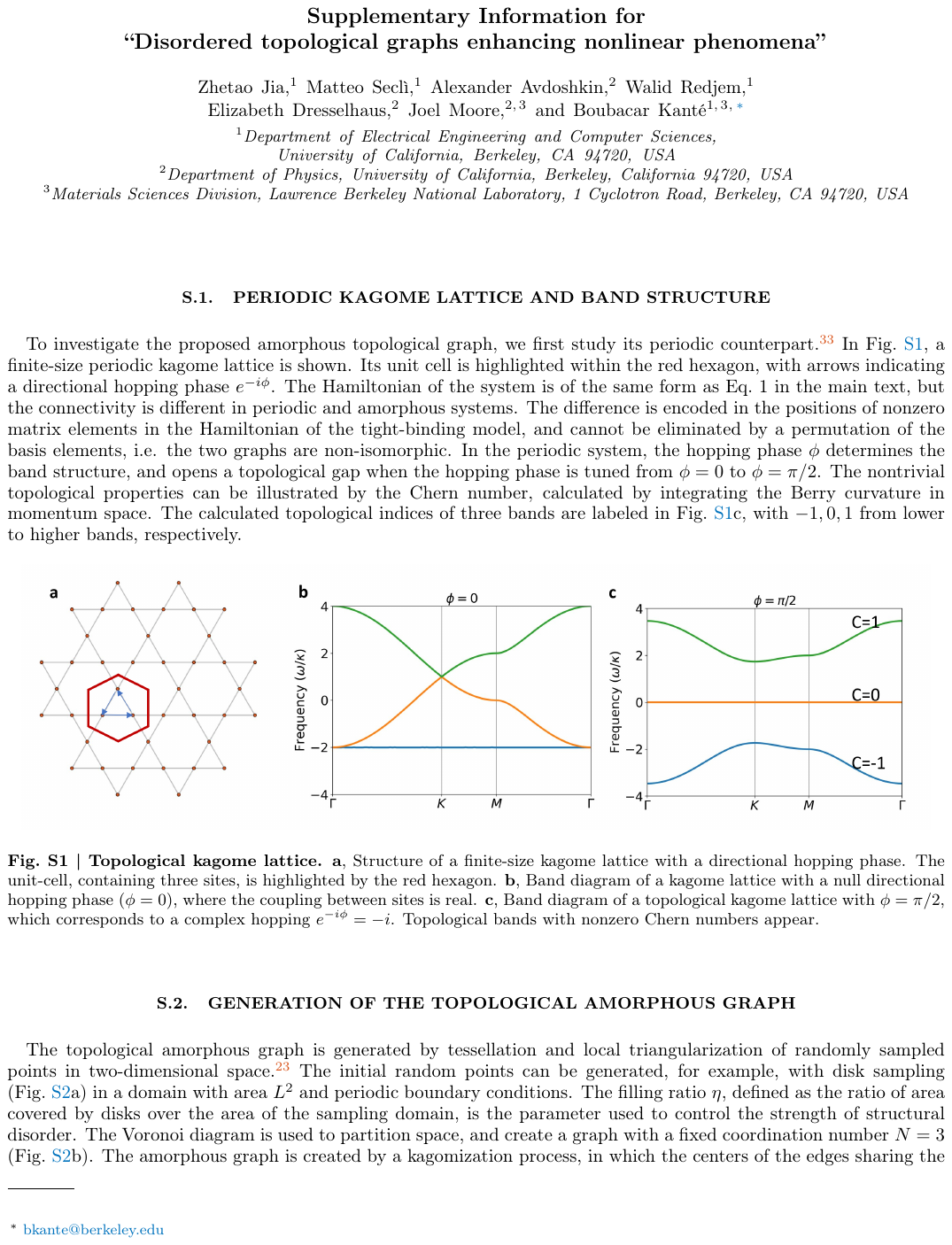}
    }
}